\documentclass{aa}  

\usepackage{graphicx}	
\usepackage{amsmath}	
\usepackage{amsmath, amsfonts, amssymb, graphics,graphicx, wrapfig, nicefrac}
\usepackage{epsfig}
\usepackage{epstopdf}
\usepackage{multirow}
\usepackage{multicol}
\usepackage{graphicx}
\usepackage{rotating}
\usepackage{mathrsfs}
\usepackage[version=4]{mhchem}
\graphicspath{{./}{Figures/}}
\usepackage{txfonts}
\begin{document}

   \title{
    A new analytic approach to infer the cosmic-ray ionization rate in hot molecular cores from HCO$^+$, N$_2$H$^+$, and CO observations}
    \titlerunning{
    A new analytic approach to infer $\zeta_2$ in hot molecular cores}

   \author{Gan Luo
          \inst{1}
          \and
          Thomas G. Bisbas\inst{2}
          \and
          Marco Padovani\inst{3}
          \and 
          Brandt A. L. Gaches\inst{4}
          }

   \institute{Institut de Radioastronomie Millimetrique, 300 rue de la Piscine, 38400, Saint-Martin d’Hères, France\\
              \email{luo@iram.fr}
        \and
            Research Center for Astronomical Computing, Zhejiang Laboratory, Hangzhou 311100, China\\
            \email{tbisbas@zhejianglab.com}
        \and
            INAF-Osservatorio Astrofisico di Arcetri, Largo E. Fermi 5, 50125 Firenze, Italy
        \and
            Department of Space, Earth and Environment, Chalmers University of Technology, Gothenburg SE-412 96, Sweden
             }

   \date{Received xx; accepted xx}

 
  \abstract
   {The cosmic-ray ionization rate ($\zeta_2$) is one of the key parameters in star formation, since it regulates the chemical and dynamical evolution of molecular clouds by ionizing molecules and determining the coupling between the magnetic field and gas.}
   {However, measurements of $\zeta_2$ in dense clouds (e.g., $n_{\rm H} \geq 10^4$\,cm$^{-3}$) are difficult and sensitive to the model assumptions. The aim is to find a convenient analytic approach that can be used in high-mass star-forming regions (HMSFRs), especially for warm gas environments such as hot molecular cores (HMCs).}
   {We propose a new analytic approach to calculate $\zeta_2$ through HCO$^+$, N$_2$H$^+$, and CO measurements. By comparing our method with various astrochemical models and with observations found
   in the literature, we identify the parameter space for which the analytic approach is applicable.}
   {Our method gives a good approximation, to within $50$\%, of $\zeta_2$ in dense and warm gas (e.g., $n_{\rm H} \geq 10^4$\,cm$^{-3}$, $T = 50, 100$\,K) for $A_{\rm V} \geq 4$\,mag and $t \geq 2\times10^4$\,yr at Solar metallicity. The analytic approach gives better results for higher densities. However, it starts to underestimate the CRIR at low metallicity ($Z = 0.1Z_\odot$) and high CRIR ($\zeta_2 \geq 3\times10^{-15}$ s$^{-1}$). By applying our method to the OMC-2 FIR4 envelope and the L1157-B1 shock region, we find $\zeta_2$ values of  $(1.0\pm0.3)\times10^{-14}$ s$^{-1}$ and $(2.2\pm0.4)\times10^{-16}$ s$^{-1}$, consistent with those previously reported.}
   {We calculate $\zeta_2$ toward a total of 82 samples in HMSFRs, finding that the average value of $\zeta_2$ toward all HMC samples ($\zeta_2$ = (7.4$\pm$5.0)$\times$10$^{-16}$\,s$^{-1}$) is more than an order of magnitude higher than the theoretical prediction of cosmic-ray attenuation models, favoring the scenario that locally accelerated cosmic rays in embedded protostars should be responsible for the observed high $\zeta_2$.}


   \maketitle
%

\section{Introduction}

Cosmic rays (CRs) play a major role in the physical and chemical evolution of dense clouds. In well-shielded regions where far-ultraviolet (FUV) photons cannot penetrate, namely where the visual extinction (A$_{\rm V}$) is larger than $3-4$~mag \citep{Wolfire2010}, CRs initiate the chemistry by ionizing various molecules and producing ions (e.g., H$_3^+$)\footnote{Assuming no other sources of ionization \citep[e.g., X-rays in the vicinity of supernova remnants,][]{Yusef-Zadeh2003,Indriolo2023}.}. On the other hand, CRs determine the degree of ionization in dense clouds and regulate the coupling between the magnetic field and gas, which, in turn, affects the dynamic timescale during the star formation process \citep{Dalgarno2006, Padovani2014, Padovani2020, Padovani2024b}. 

Ions (e.g., H$_3^+$, OH$^+$ and H$_2$O$^+$) are the most favored tracers to constrain the cosmic-ray ionization rate (CRIR), since the formation and destruction of these species are relatively simple and allow us to estimate almost directly the CRIR \citep{Gerin2010,Indriolo2012,Indriolo2015,Neufeld2017}. In dense clouds, such methodology is, however, not available due to the large extinction and low abundance of the above-mentioned ions. 

The abundance ratios of DCO$^+$/HCO$^+$ and HCO$^+$/CO have been proposed to infer the ionization fraction and CRIR in cold prestellar cores \citep{Caselli1998}. However, the deuteration ratios are sensitive to the initial ortho-to-para ratio, the evolution timescale, and the source physical conditions \citep[e.g., gas density and temperature,][]{Shingledecker2016}. Thus, the resultant CRIR from the above method could be overestimated by several orders of magnitude \citep{Sabatini2023,Redaelli2024}. More recently, an analytic approach using otho-H$_2$D$^+$ and CO was proposed to estimate the CRIR \citep{Bovino2020} in cold prestellar cores where depletion of molecules is crucial. In addition, rovibrational H$_2$ lines could be a potentially useful method to derive CRIR value in the cold molecular cloud \citep{Bialy2020,Padovani2022,Bialy2022,Gaches2022b}, which would be, in principle, detected by the James Webb Space Telescope (JWST). 

In high-mass star-forming regions (HMSFRs), the gas temperature increases in the more evolved phase \citep[e.g., $T \sim 100$\,K in hot molecular cores, HMCs,][]{Blake1987,Beuther2007b,Luo2019,Qin2022}, depletion and deuteration are inefficient, so that the methods described above cannot be adopted. Over the past decade, different molecular line tracers combined with astrochemical models have been used to probe the CRIR in HMSFRs, such as HCO$^+$ and N$_2$H$^+$ \citep{Ceccarelli2014b,Redaelli2021}, and carbon-chain molecules \citep[e.g., HC$_3$N, HC$_5$N,][]{Fontani2017,Favre2018}. While chemical modeling requires large computational resources, a convenient analytic approach would be useful in determining the variance of CRIR with the spatial distribution and evolution stage in HMSFRs.

In this work, we propose a new analytic approach to estimate the CRIR in HMSFRs, especially for high-mass protostellar objects (HMPOs) and HMCs. HMPOs are in a more evolved phase than the cold infrared dark cloud (IRDC). They consist of massive accreting protostars (e.g, $M > 8 M_\odot$) and exhibit bright mid-infrared emission and strong outflows \citep{Beuther2002b,Beuther2007a,Zinnecker2007,Motte2018}. The characteristic size of HMPOs is $\sim$0.5\,pc, and the average density is between $10^4$ and $10^6$\,cm$^{-3}$ \citep{Beuther2002a}. 
HMCs appear in the later stage of HMPOs and are characterized by high gas temperatures (e.g., $T \gtrsim$ 100\,K), densities ($n_{\rm H} \gtrsim 10^6$\,cm$^{-3}$), and rich molecular lines \citep{Blake1987,Kurtz2000,Beuther2007b,Qin2022}. 
We compare our approach with chemical simulations from {\sc 3d-pdr} and {\sc uclchem} and demonstrate that our method can give a good estimation of CRIR for HMPOs and HMCs. 

The paper is organized as follows. We present the chemical analysis and describe the analytic approach in Sect.~\ref{sec:methods}. In Sect.~\ref{sec:results}, we compare our approach with chemical simulations. We apply our method to two well-known sources and compare our results with previous modeling in the literature in Sect.~\ref{sec:discussion}. The uncertainties and robustness of the approach are also discussed. In Sect.~\ref{sec:crir in msfr}, we calculate the CRIR toward a large sample of HMSFRs. The main results and conclusions are summarised in Sect.~\ref{sec:conclusions}.

\section{Methods}\label{sec:methods}

\subsection{Chemical analysis}\label{sec:chem}

We describe, below, the formation and destruction reactions of H$_3^+$, HCO$^+$, N$_2$H$^+$, and H$_3$O$^+$ in dense and warm clouds (e.g., $n_{\rm H} \geq 10^4$\,cm$^{-3}$, $T \geq 50$\,K), where the depletion is negligible. 

The formation of H$_3^+$ in dense clouds starts from the ionization of H$_2$ by CRs \citep[see, e.g.,][]{Glassgold1973}:
\begin{align}
\rm H_2 + CR  \rightarrow &\  \rm H_2^+ + e^- \tag{R1},\label{r:R1}\\ 
\rm  H_2^+ + H_2 \rightarrow &\ \rm H_3^+ + H \tag{R2}.\label{r:R2}
\end{align}
Then, H$_3^+$ can be removed by the most abundant neutral species \citep{Indriolo2012}:
\begin{align}
\rm H_3^+ + CO  \rightarrow &\  \rm HCO^+ + H_2 \notag,\\ 
\rm   \rightarrow &\  \rm HOC^+ + H_2 \tag{R3},\label{r:R3}\\ 
\rm  H_3^+ + N_2 \rightarrow &\ \rm N_2H^+ + H_2 \tag{R4}.\label{r:R4}\\
\rm  H_3^+ + O \rightarrow &\ \rm OH^+ + H_2 \notag,\\
\rm   \rightarrow &\ \rm H_2O^+ + H_2 \tag{R5},\label{r:R5}
\end{align}
and electrons:
\begin{align}
\rm H_3^+ + e^-  \rightarrow &\  \rm H_2 + H \notag,\\
\rm   \rightarrow &\  \rm H + H + H \tag{R6}.\label{r:R6}
\end{align}
Reactions \ref{r:R3} and \ref{r:R4} are the main formation channels of HCO$^+$\footnote{Reaction \ref{r:R8} below also contribute to the formation of HCO$^+$ but with a minor role.} and N$_2$H$^+$, respectively. 

The corresponding destruction channels of HCO$^+$ and N$_2$H$^+$ are \citep{Dalgarno2006}:
\begin{align}
\rm HCO^+ + e^-  \rightarrow &\  \rm CO + H \tag{R7},\label{r:R7}
\end{align}
and\footnote{The reaction \ref{r:R10} is endothermic by 0.3 eV, thus, it could only occur under specific conditions (e.g., shocked or turbulent environments). Considering the small reaction rate of \ref{r:R10} (Table~\ref{table:rates}), ignoring it would only contribute a marginal decrease ($\sim$10\%) to the calculated $\zeta_2$ using the analytic approach.}
\begin{align}
\rm N_2H^+ + CO  \rightarrow &\  \rm HCO^+ + N_2 \tag{R8},\label{r:R8}\\
\rm N_2H^+ + e^-  \rightarrow &\  \rm H + N_2 \tag{R9}.\label{r:R9}\\
\rm N_2H^+ + O  \rightarrow &\  \rm OH^+ + N_2 \tag{R10}.\label{r:R10}
\end{align}

Considering reactions \ref{r:R1}-\ref{r:R6}, CRIR ($\zeta_2$, referring to the ionization rate of H$_2$) can be written as:
\begin{align}
\rm \zeta_2 =&\  \rm \frac{{\it n}(H_3^+)}{{\it n}(H_2)}\left[{\it n}(CO)k_{R3}+{\it n}(N_2)k_{R4}+{\it n}(O)k_{R5}+{\it n}(e^-)k_{R6}\right].
\label{eq:eq1}
\end{align}
Similarly, from reactions \ref{r:R4} and \ref{r:R8}--\ref{r:R10} we can write:
\begin{align}
\rm {\it n}(H_3^+){\it n}(N_2)k_{R4} =&\ \rm {\it n}(N_2H^+)\left[{\it n}(CO)k_{R8} + {\it n}(e^-)k_{R9} + {\it n}(O)k_{R10}\right].
\label{eq:eq3}
\end{align}

Considering the charge balance, the number density of electrons should be equal to the most abundant ions (see Appendix \ref{sec:a1}):
\begin{align}
\rm {\it n}(e^-) =&\  \rm {\it n}(HCO^+)+{\it n}(N_2H^+)+{\it n}(H_3^+)+{\it n}(H_3O^+).
\label{eq:eq4}
\end{align}
In dense clouds, the formation of H$_3$O$^+$ starts from reaction \ref{r:R5}.
Then, it hydrogenates to form H$_2$O$^+$ and H$_3$O$^+$ \citep{Bialy2015,Indriolo2015,Luo2023b}:
\begin{align}
\rm OH^+ + H_2  \rightarrow &\ \rm H_2O^+ + H  \tag{R11}, \label{r:R11}\\ 
\rm  H_2O^+ + H_2 \rightarrow &\  \rm H_3O^+ + H \tag{R12}.
\label{r:R12}
\end{align}

Electron recombination reactions dominate the destruction of H$_3$O$^+$:
\begin{align}
\rm H_3O^+ + e^-  \rightarrow &\ \rm OH + H + H \notag, \\ 
\rm   \rightarrow &\  \rm OH + H_2  \tag{R13}, \label{r:R13}\\
\rm   \rightarrow &\  \rm H_2O + H \notag.
\end{align}
From reactions \ref{r:R5}, \ref{r:R10}, and \ref{r:R11}-\ref{r:R13}, $n({\rm H_3O^+})$ can be written as:
\begin{align}
\rm {\it n}(H_3O^+) =&\  \rm \frac{{\it n}(O){\it n}(H_3^+)k_{R5}+{\it n}(N_2H^+){\it n}(O)k_{R10}}{{\it n}(e^-)k_{R13}}.
\label{eq:eq5}
\end{align}
Combining Eqs.~(\ref{eq:eq4}) and (\ref{eq:eq5}), $n{\rm (H_3^+)}$ can be written as:
\begin{align}
\rm {\it n}(H_3^+) =&\  \rm \frac{{\it n}(e^-)^2-{\it n}(e^-)\left[{\it n}(HCO^+)+{\it n}(N_2H^+)\right]-\alpha}{{\it n}(e^-)+{\it n}(O)k_{R5}/k_{R13}},
\label{eq:eq6}
\end{align}
where $\alpha$ is
\begin{align}
\rm \alpha =&\  \rm \frac{{\it n}(N_2H^+){\it n}(O)k_{R10}}{k_{R13}},
\end{align}
Then, $n{\rm (e^-)}$ can be calculated from Eqs.~(\ref{eq:eq3}) and (\ref{eq:eq6}):
\begin{align}
\rm {\it n}(e^-) =&\  \rm \frac{\beta+\sqrt{\beta^2+4\gamma\theta}}{2\gamma},
\label{eq:eq7}
\end{align}
where 
\begin{align}
\rm \beta =&\  \rm \frac{{\it n}(N_2H^+)}{{\it n}(N_2)k_{R4}}\left[\frac{{\it n}(O)k_{R5}k_{R9}}{k_{R13}}+{\it n}(CO)k_{R8}+{\it n}(O)k_{R10}\right] \notag \\
&\ \rm +{\it n}(HCO^+)+{\it n}(N_2H^+),
\end{align}
\begin{align}
\rm \gamma =&\  \rm 1-\frac{{\it n}(N_2H^+)k_{R9}}{{\it n}(N_2)k_{R4}},
\end{align}
and 
\begin{align}
\rm \theta =&\  \rm \alpha\left[1+k_{R5}\frac{{\it n}(CO)k_{R8}+{\it n}(O)k_{R10}}{{\it n}(N_2)k_{R4}k_{R10}}\right].
\end{align}

\subsubsection{The new analytic approach}
\label{sssec:analytic}
To determine $\zeta_2$ through our new methodology it is needed:
\begin{enumerate}
    \item a proper constraint on $n{\rm (H_2)}$;
    \item to determine the abundances of HCO$^+$, N$_2$H$^+$ and CO from observations;
    \item an assumption of the initial abundances of N$_2$ and O.
\end{enumerate}

In dense cores where hydrogen is mostly in molecular form, we can divide all number density terms ({\it n}(x)) by {\it n}(H$_2$) in Eqs.~(\ref{eq:eq1}) to (\ref{eq:eq7}). Thus, the calculated CRIR ($\widetilde{\zeta_2}$, we distinguish between the calculated CRIR ($\widetilde{\zeta_2}$) and the input one ($\zeta_2$) from chemical simulations) can be written as:
\begin{align}
\rm \widetilde{\zeta_2} =&\  \rm {\it n}(H_2){\it f}(H_3^+)\left[{\it f}(CO)k_{R3}+{\it f}(N_2)k_{R4}+{\it f}(O)k_{R5}+{\it f}(e^-)k_{R6}\right],
\label{eq:eq8}
\end{align}
where $f$(x) is the relative abundance of species ``x'' with respect to H$_2$, {\it n}(H$_2$) is the number density of H$_2$, and $f$(e$^-$) can be obtained from Eq.~\ref{eq:eq7}.

For the calculations of $\widetilde{\zeta_2}$, we use the latest reaction rates from the UMIST database \citep[hereafter, UMIST2022,][]{Millar2024} unless otherwise stated.

\subsection{Chemical models}\label{sec:chem models}

To explore the behavior of our analytic approach, we employ the publicly available astrochemical codes\footnote{https://uclchem.github.io/} {\sc 3d-pdr} \citep{Bisbas2012} and {\sc uclchem} \citep{Holdship2017}. {\sc 3d-dpr} is a photodissociation region code that models one- and three-dimensional density distributions. It calculates the attenuation of the FUV radiation field in every depth point and outputs the self-consistent solutions of chemical abundances, gas and dust temperature, emissivities, and level populations by performing iterations over thermal balance. In the {\sc 3d-pdr} models, we consider more than 200 species with 3000 reactions taken from the UMIST2012\footnote{We note that the reaction rates of \ref{r:R5}, \ref{r:R8}, \ref{r:R11}, and \ref{r:R12} have changed by up to a factor of $\sim$3 between the UMIST2012 and UMIST2022. In that case, we calculate $\widetilde{\zeta_2}$ in Sect.~\ref{sec:3dpdr} using the reaction rates from UMIST2012 only when comparing with {\sc 3d-pdr} simulations.} \citep{McElroy2013}. The initial gas phase abundances of different elements are listed in Table \ref{tab:abundances}. We note that the initial element abundances are not expected to be the same everywhere. However, as long as the abundances of key species do not change significantly, the assumptions and conclusions in Sect.~\ref{sec:chem} remain valid (see \S\ref{sec:variation of z} for more discussion on different metallicities.).
We also consider the treatment of CR attenuation incorporated into {\sc 3d-pdr}, as presented in \citet{Gaches2019,Gaches2022a}. Unless otherwise stated, all {\sc 3d-pdr} models presented here are calculated until the thermal balance is reached.

We perform additional simulations with {\sc uclchem} \citep{Holdship2017} to test how the gas-grain reactions and depletion processes influence the results. {\sc uclchem} is a time-dependent astrochemical code focusing on gas-grain reactions. It considers freeze-out processes, thermal, and non-thermal desorption processes and outputs the chemical abundances at a given temperature, density, and $A_{\rm V}$. Contrary to {\sc 3d-pdr} calculations where the code terminates as long as the thermal balance is reached, the simulations in {\sc uclchem} are isothermal. We set three temperatures in these simulations, $T_{\rm k}$ = 15\,K, 50\,K, and 100\,K. 15 K is the typical temperature of the cold dark cloud and it is below the temperature for CO desorption ($T \sim 20$ K) \citep{Oberg2011}. 50 K is above the sublimation temperature of simple molecules (e.g., CO) but below the desorption temperature of water ice and COMs. 100 K is the typical temperature of HMCs and most of the species would be released from the ice phase \citep{vanDishoeck2014}.

In {\sc uclchem} simulations, the deuterium chemical network of \citet{Majumdar2017} is adopted. To reduce the computational expense, species with molecular weight higher than 66 were excluded (mostly carbon-chain molecules and complex organic molecules that contain more than 5 carbon atoms). Excluding these molecules may not greatly impact the results since they gradually build up the abundances from C and C$^+$, and have little chemical connection with the simple species we mentioned above. Besides, their abundances are usually too low \citep[e.g., 10$^{-10}$ and below,][]{Herbst2009} to impact the simple species. The chemical network contains a total of $\sim800$ species and $\sim 3\times10^4$ reactions. The initial gas-phase abundance of HD is adopted as the value measured in the Local Bubble \citep[1.6$\times$10$^{-5}$,][]{Linsky2006}\footnote{The measured D/H ratios in high column density regions are lower, presumably due to the depletion of D on dust grain \citep{Linsky2006,Friedman2023}. However, the D/H ratio only matters in our analysis for the dark cloud where depletion of molecules is significant (See Sect.~\ref{sec:uclchem}).}. The remaining element abundances are the same as in Table \ref{tab:abundances}. For all the simulations in both {\sc 3d-pdr} and {\sc uclchem}, the cloud radius is set as 1\,pc.

\begin{table}
\caption{Initial gas-phase element abundances used in our simulations.}
    \centering
    \begin{tabular}{ccc}
    \hline\hline
       Elements  & Abundance relative to H & References \\
       \hline
       H  & 5$\times$10$^{-1}$ & / \\
       H$_2$  & 2.5$\times$10$^{-1}$ & /\\
       He  &  9$\times$10$^{-2}$ & a \\
       C$^+$ & 1.5$\times$10$^{-4}$ & b,c \\
       O & 3$\times$10$^{-4}$ & d,e \\
       N & 7.6$\times$10$^{-5}$ & f \\
       S$^+$ & 8$\times$10$^{-8}$ & c, g \\
       Si$^+$ & 8$\times$10$^{-9}$ & c, g \\
       Mg$^+$ & 7$\times$10$^{-9}$ & c, g \\
       Fe$^+$ & 3$\times$10$^{-9}$ & c, g \\
       Na$^+$ & 2$\times$10$^{-9}$ & c, g \\
       P$^+$  & 2$\times$10$^{-10}$ & c, g \\
       Cl$^+$ & 1$\times$10$^{-9}$ & c, g \\
       F$^+$ & 6.68$\times$10$^{-9}$ & h \\
       \hline
    \end{tabular}
    \tablefoot{References: (a) \citet{Wakelam2008}. (b) \citet{Sofia1997}. (c) \citet{Hincelin2011}. (d) \citet{Meyer1998}. (e) \citet{Lis2023}. (f) \citet{Meyer1997}. (g) \citet{Majumdar2017}. (h) \citet{Neufeld2005}. }
    \label{tab:abundances}
\end{table}

\section{Results}\label{sec:results}
\subsection{Comparison to {\sc 3d-pdr} simulations}\label{sec:3dpdr}

\begin{figure}
	\includegraphics[width=1.0\columnwidth]{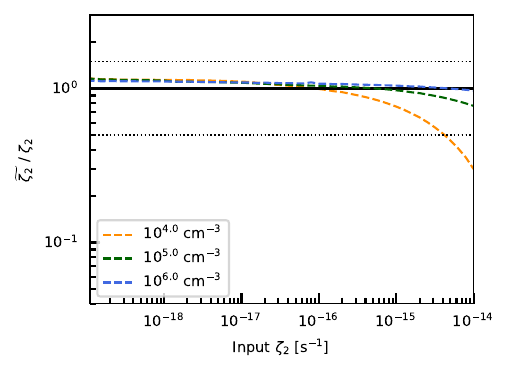}
    \caption{The comparison between our approach and {\sc 3d-pdr} simulations. The x-axis denotes the input CRIR of each simulation, where the CRIR in each simulation is treated as uniform. The y-axis denotes the ratio between the calculated CRIR and the input one ($\widetilde{\zeta_2}$/$\zeta_2$) at $A_{\rm V}$ = 20\,mag. The black horizontal line represents the fiducial value, and the horizontal dotted lines represent 50\% of the deviation from the input values. Orange, green, and blue dashed lines represent results at $n_{\rm H}$ = 10$^4$, 10$^5$, and 10$^6$\,cm$^{-3}$. The external FUV intensity for all simulations is normalized to the Draine unit ($\chi/\chi_0$ = 1).}
    \label{fig:fig1}
\end{figure}

We test three uniform density clouds ($n_{\rm H}$ = 10$^4$, 10$^5$, and 10$^6$\,cm$^{-3}$) with different uniform CRIR ($10^{-19} \le \zeta_2/{\rm s}^{-1} \le 10^{-14}$) input. Figure \ref{fig:fig1} shows the ratio between CRIR derived from Eq.~\ref{eq:eq8} and the input value ($\widetilde{\zeta_2}$/$\zeta_2$) at $A_{\rm V}$ = 20\,mag. In most of the simulations, the calculated CRIR from the analytic approach is within 50\% of the model input. The analytic approach underestimates the input CRIR by up to a factor of 3 at $n_{\rm H}$ = 10$^4$ cm$^{-3}$ and $\zeta_2 = 10^{-14}$ s$^{-1}$. This is due to the destruction of CO molecules under high CRIR \citep{Bialy2015,Bisbas2015,Bisbas2017}, thus, the destruction pathways we analyzed in Sect. \ref{sec:chem} change. As the density increases, the analytic approach gives a more accurate result.

\begin{figure*}
	\includegraphics[width=2.0\columnwidth]{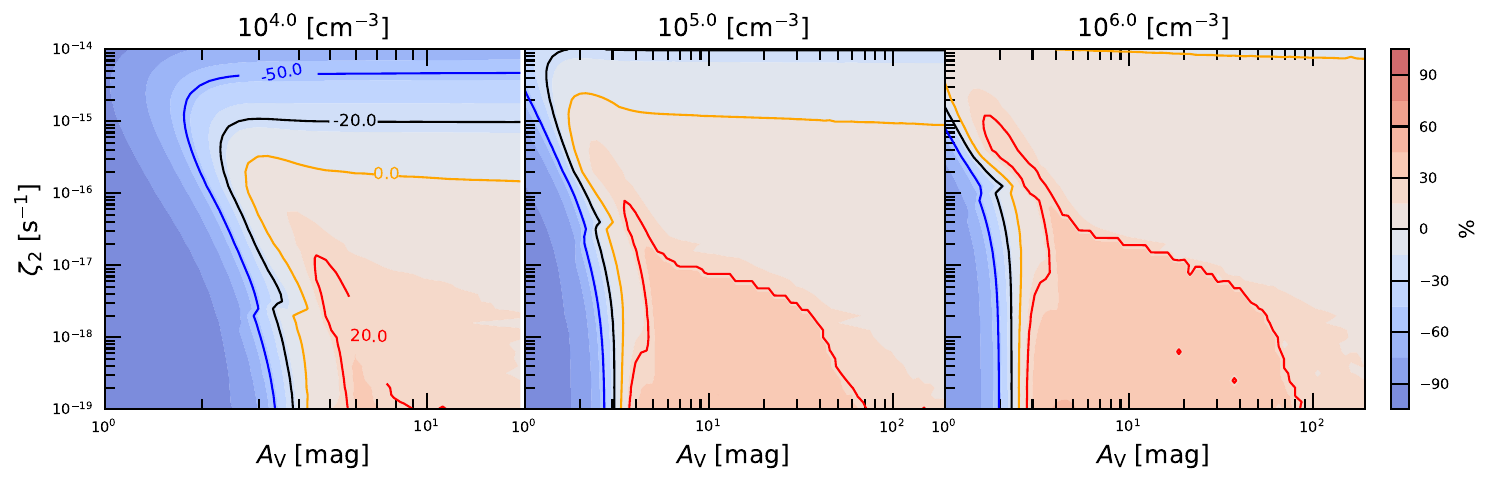}
    \caption{The deviation maps between the analytic approaches and the input values at different $A_{\rm V}$ and $\zeta_2$ for $n_{\rm H}$ = 10$^4$ (left), 10$^5$ (middle), and 10$^6$\,cm$^{-3}$ (right). Red, orange, black, and blue curves denote the 20\%, 0\%, -20\%, and -50\% deviations.}
    \label{fig:fig2}
\end{figure*}

Figure \ref{fig:fig2} shows the deviation maps between the analytic approach and the input values at different $A_{\rm V}$ and $\zeta_2$. Our analytic approach gives a good approximation, to within $50\%$,
if $\zeta_2 \leq 5\times10^{-15}$\,s$^{-1}$ and $A_{\rm V} \geq 4$\,mag at $n_{\rm H}$ = 10$^4$\,cm$^{-3}$, and $A_{\rm V} \geq 4$\,mag for any $\zeta_2$ at $n_{\rm H}$ = 10$^5$\,cm$^{-3}$ and $n_{\rm H}$ = 10$^6$\,cm$^{-3}$.

\begin{figure}
	\includegraphics[width=1.0\columnwidth]{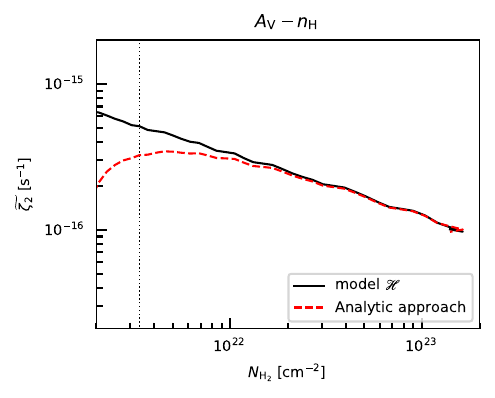}
    \caption{Comparison of the calculated CRIR from the analytic approach ($\widetilde{\zeta_2}$, red dashed curve) and the CR attenuation model (black solid curve). The vertical dashed line represents the 50\% deviation between the analytic approach and the CR attenuation model.}
    \label{fig:fig3}
\end{figure}

In addition, we test the CR attenuation model with a variable density distribution, in which the CR spectrum is adopted as the high-energy spectrum model $\mathscr{H}$ from \citet{Ivlev2015}, and the density distribution of the cloud adopts the $A_{\rm V}$-$n_{\rm H}$ relation of \citet{Bisbas2023}. This relation has been found to reproduce reasonably well and at a minimal computational cost, the results from computationally expensive three-dimensional astrochemical models.   
Figure \ref{fig:fig3} shows the comparison between the input $\zeta_2$ and the analytical approach. At lower column densities ($N_{\rm H_2} < 3.4\times10^{21}$\,cm$^{-2}$, corresponding to $n_{\rm H} < 3\times10^3$\,cm$^{-3}$), our approach underestimates $\zeta_2$ by over 50\%. At higher column densities, our approach gives a good approximation.

\subsection{Comparison to {\sc uclchem} simulations}\label{sec:uclchem}

\begin{figure*}
	\includegraphics[width=2.0\columnwidth]{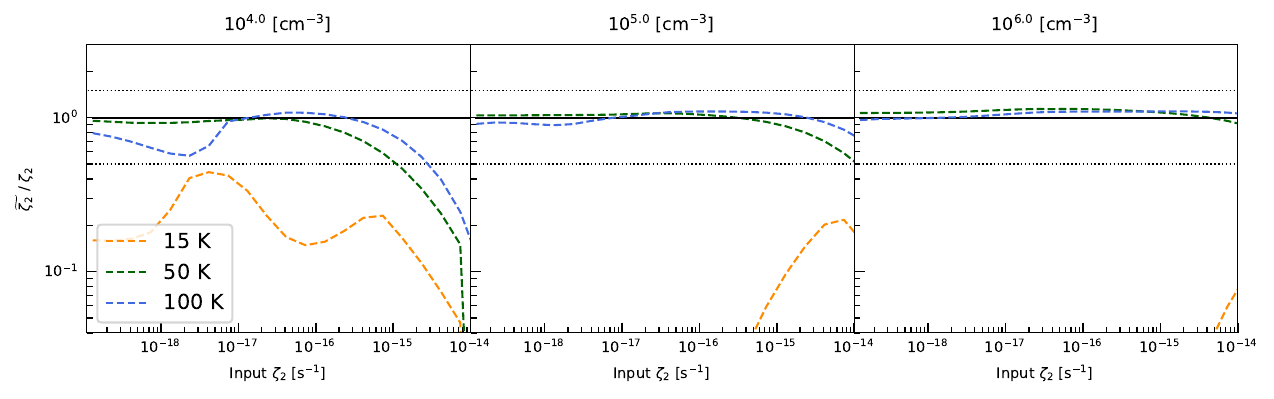}
    \caption{Comparison between our approach and the {\sc uclchem} simulations. The x-axis denotes the input CRIR of each simulation, where the CRIR in each simulation is treated as uniform. The y-axis denotes the ratio between the calculated CRIR and the input one ($\widetilde{\zeta_2}$/$\zeta_2$) at $t$ = 10$^6$ yr. The black horizontal line represents the fiducial value, and the horizontal dotted lines represent 50\% of the deviation from the input values. Orange, green, and blue dashed lines represent results at $T_{\rm k}$ = 15, 50, and 100 K. Columns from left to right represent simulations with the cloud densities $n_{\rm H}$ = 10$^4$, 10$^5$, and 10$^6$\,cm$^{-3}$. The large deviation at 15 K is due to the freeze-out processes that greatly reduce the gas-phase abundances of key species such as CO, thus, the chemical reactions we described in Sect.~\ref{sec:chem} are no longer satisfied.}
    \label{fig:fig4}
\end{figure*}

Figure \ref{fig:fig4} shows the comparison between the analytic approach with {\sc uclchem} simulations for different temperatures and densities. At low temperatures ($T_{\rm k}$ = 15\,K), most of the species (including CO and O) would freeze out onto dust grain, the analytic approach underestimates the CRIR by a factor from a few ($n_{\rm H}$ = 10$^4$\,cm$^{-3}$) to several orders of magnitude (at $n_{\rm H}$ = 10$^5$ and 10$^6$\,cm$^{-3}$). This is expected since the depletion processes greatly reduce the abundances of CO, O, and N$_2$ by several orders of magnitude such that the destruction pathways of H$_3^+$ we considered in Sect.~\ref{sec:methods} (reactions \ref{r:R3}-\ref{r:R5}) are no longer the dominant H$_3^+$ destruction mechanism. Instead, deuteration fractionation cannot be ignored such that the destruction of H$_3^+$ and the reaction channels should involve deuterated species \citep{Millar1989,Ceccarelli2014a,Caselli2019}.

When the gas temperature is higher than the sublimation temperature of CO (the cases of $T_{\rm k}$ = 50 and 100\,K), the analytic approach gives a better estimation, to within 50\% for most cases. The higher the gas temperature, the better estimation is given by the analytic approach. Our analytic approach gives a more accurate estimation for higher density simulations, the same as we found with the {\sc 3d-pdr} simulations.

\begin{figure*}
	\includegraphics[width=2.0\columnwidth]{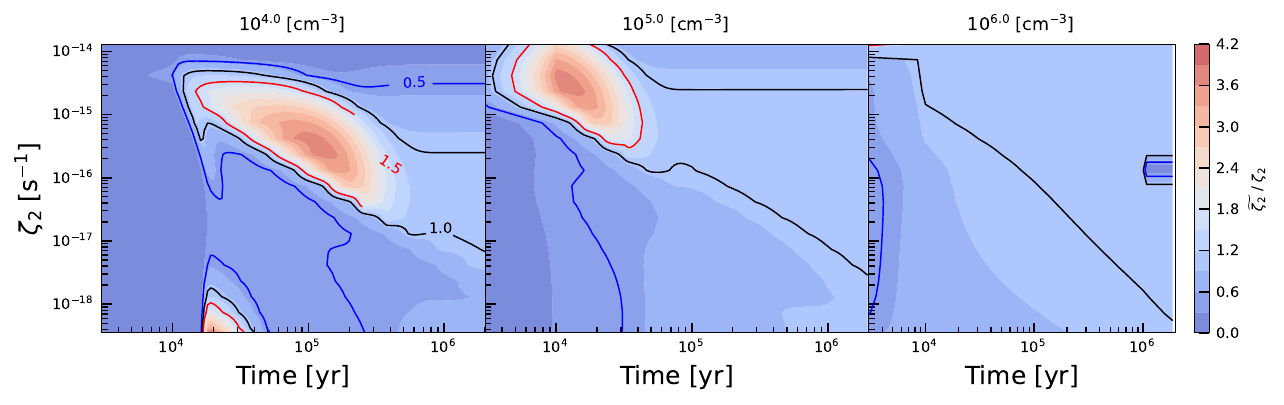}
    \caption{The deviation between the analytic approaches and the input values at different evolution timescales and $\zeta_2$. Red, black, and blue curves denote the 50\%, 0\%, and -50\% deviations.}
    \label{fig:fig5}
\end{figure*}

Figure \ref{fig:fig5} shows the deviation map between the analytic approach and the input values at different evolution timescales and $\zeta_2$. For early timescales, where the molecules are still building their abundances (e.g., at $n_{\rm H}$ = 10$^4$\,cm$^{-3}$ and $t \leq 10^4$\,yr), the analytic approach underestimates the CRIR. At $n_{\rm H}$ = 10$^4$\,cm$^{-3}$ and $t \geq 2\times10^4$ yr, the analytic approach gives a reasonable $\zeta_2$ estimation except for $t \approx 10^5$ yr and $\zeta_2 \approx 5\times10^{-16}$ s$^{-1}$, where the calculated value is overestimated by a factor of 3.7. For $n_{\rm H}$ = 10$^5$\,cm$^{-3}$, the analytic approach gives a good approximation of the CRIR for the majority of the parameter space if $t > 10^4$ yr. The maximum deviation between the analytic approach and the input value ($\widetilde{\zeta_2}$/$\zeta_2$ = 2.6) appears at $t \approx 10^4$ yr and $\zeta_2 \approx 5\times10^{-15}$ s$^{-1}$. At $n_{\rm H}$ = 10$^6$\,cm$^{-3}$, the analytic approach gives a good estimate for all the parameter space we modeled. 
The results suggest that the analytic approach can be used for conditions of warm dense gas, namely in HMPOs and HMCs. 

\section{Discussion}\label{sec:discussion}

\subsection{Robustness of the analytic approach}\label{sec:robustness}

In the above simulations, we ignore the variation of initial element abundances (e.g., in different metallicity environments), the size of the source, and the gas density and temperature during the evolution processes. In the following sections, we discuss the robustness of the analytic approach under different cases. 

\subsubsection{Variation of metallicity}\label{sec:variation of z}

\begin{figure}
	\includegraphics[width=1.0\columnwidth]{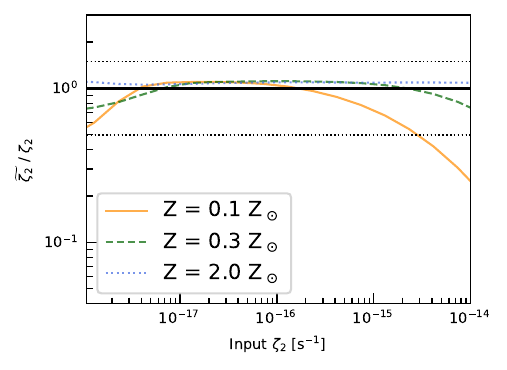}
    \caption{The comparison between our approach and {\sc uclchem} simulations under different metallicities ($Z$). The temperature and density are adopted as 100 K and $10^6$ cm$^{-3}$. The solid, dash, and dotted lines represent $Z$ = 0.1, 0.3, and 2 $Z_{\odot}$. The black horizontal line represents the fiducial value, and the horizontal dotted lines represent 50\% of the deviation from the input values.}
    \label{fig:figz}
\end{figure}

The variation of metallicity would change the initial abundances of heavy elements (heavier than He). We test three metallicities ($Z$ = 0.1, 0.3, and 2 $Z_{\odot}$, which correspond to low-metallicity, outer Galaxy/Large Magellanic Cloud (LMC), and the Galactic Center) with {\sc uclchem} at $n_{\rm H}$ = 10$^6$ cm$^{-3}$. The results are shown in Fig.~\ref{fig:figz}. The analytic approach gives a good estimation of $\zeta_2$ at $Z$ = 0.3 and 2 $Z_{\odot}$. However, the analytic approach starts to underestimate the value of $\zeta_2$ at $Z = 0.1 Z_{\odot}$ when the CRIR is high ($\zeta_2 \geq 3\times10^{-15}$ s$^{-1}$). At $Z = 0.1 Z_{\odot}$, the abundances of heavy elements are reduced by a factor of 10, while the abundances of D and He remain the same. Furthermore, due to the high CRIR, key species like CO would be easily destroyed, leading to comparable or lower abundances than that of HD. Thus, the main destruction channels of H$_3^+$ we considered in Sect.~\ref{sec:methods} would change. The results suggest that our analytic approach may only give a lower limit for metal-poor starburst galaxies where the CRIR is supposed to be extremely high.

\subsubsection{Variation of source size}\label{sec:variation of r}

\begin{figure}
	\includegraphics[width=1.0\columnwidth]{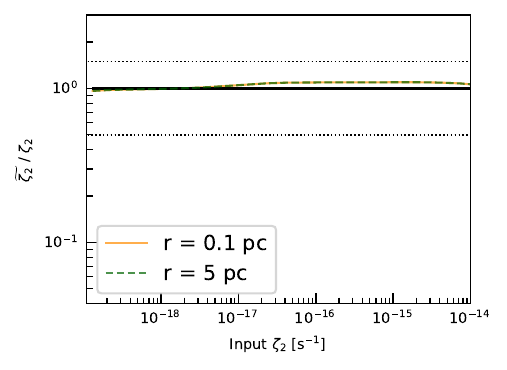}
    \caption{The comparison between our approach and {\sc uclchem} simulations under different source size (r = 0.1 and 5 pc). The temperature and density are adopted as 100 K and $10^6$ cm$^{-3}$. The solid and dash lines represent r = 0.1 and 5 pc, respectively. The black horizontal line represents the fiducial value, and the horizontal dotted lines represent 50\% of the deviation from the input values.}
    \label{fig:fig_size}
\end{figure}

In our simulations, we set the cloud radius to be 1 pc. In reality, the size of the source could change significantly in different environments. Figure \ref{fig:fig2} shows the deviation maps from {\sc 3d-pdr} simulations, where the analytic approach gives a good estimation of CRIR for $A_{\rm V}\gtrsim4\,{\rm mag}$. This value corresponds to $0.25\,{\rm pc}$ at $n_{\rm H} = 10^4$ cm$^{-3}$ and $2.5\times10^{-3}\,{\rm pc}$ ($\sim500\,{\rm AU}$) at $n_{\rm H} = 10^6$ cm$^{-3}$, suggesting a minimal source size that the analytic approach can give a good CRIR estimation. We also run simulations with two different source sizes (r = 0.1, 5 pc) in {\sc uclhem}. As seen in Fig.~\ref{fig:fig_size}, changing the source size has no impact on the conclusions.

\subsubsection{HMC models}\label{sec:hmc models}
To test the robustness of our analytic approach, we perform tests with more physical HMC models, which consider both the variation of density and temperature with evolution timescale. Each of the HMC models consists of two stages. In the first stage, we perform simulations of the isothermal free-fall collapse (representing a prestellar phase) with an initial temperature of 10\,K. The free-fall collapse phase starts from a cloud with $n_{\rm H} = 10^3$ cm$^{-3}$, reaching higher densities ($n_{\rm H}^{\rm final}$ = 10$^4$, 10$^5$, and 10$^6$\,cm$^{-3}$) at a free-fall timescale\footnote{The choice of the initial density would not influence the conclusion but could greatly reduce the time cost of each simulation.} \citep[see][for a detailed description of the density profiles in collapse models]{Priestley2018}. In the second stage, we perform simulations with a gradually increasing gas temperature (representing a formed HMPO or HMC) due to the heating of a central massive star. The gas densities and initial composition of all the species in the second stage inherit from the final state of the free-fall collapse. The mass of the central star in the HMC model is set to 10\,M$_\odot$, and the gas temperature increases from 10\,K to 200\,K within a few 10$^5$\,yr (see \citealt{Viti2004} and \citealt{Awad2010} for more details of the treatment of temperature).

\begin{figure}
	\includegraphics[width=1.0\columnwidth]{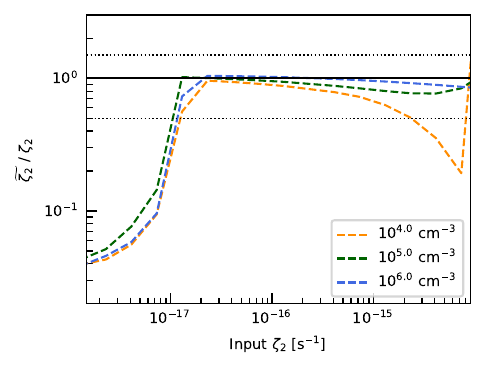}
    \caption{Comparison of $\widetilde{\zeta_2}/\zeta_2$ as a function of the input $\zeta_2$ in the HMC models. Orange, green, and blue lines represent results at $n_{\rm H} = 10^4, 10^5, 10^6$\,cm$^{-3}$. Dashed horizontal lines represent the 50\% deviation from the input values.}
    \label{fig:fig7}
\end{figure}

Figure \ref{fig:fig7} shows the comparison between the analytic approach and the HMC models. The analytic approach gives a reasonable estimation for all models at $\zeta_2 \geq 10^{-17}$\,s$^{-1}$, while it underestimates $\zeta_2$ by up to a factor of 25 at $\zeta_2 \approx 10^{-18}$\,s$^{-1}$. As shown in Fig.~\ref{fig:a1} this is due to an underestimation of electron abundance through Eq.~\ref{eq:eq3} at very low CRIR, where the formation of ions (e.g., HCO$^+$ and N$_2$H$^+$) is suppressed while Eq.~\ref{eq:eq3} does not consider the contribution from other ions such as C$^+$, Mg$^+$, etc. 

\subsection{Uncertainties}\label{sec:uncertainty}

There are a few factors that may bring uncertainties when using the analytic approach to calculate the CRIR. The most uncertain factors are the assumption of the gas phase abundances of O and N$_2$, since they are almost undetectable in dense clouds. On the other hand, when there is a lack of measurements of CO isotopologues, a CO abundance of 10$^{-4}$ (without depletion) is frequently assumed. In this section, we discuss the uncertainties induced by these factors.

Without depletion, most of the carbon is found in CO molecules in dense molecular clouds. Thus, CO has been widely used as an indicator of H$_2$ in large surveys \citep{Dame2001,Su2019}. The canonical abundance of CO in well-shielded molecular clouds (e.g., $A_{\rm V} > 4$\,mag) is $\approx 10^{-4}$ \citep{Frerking1982,Pineda2010,Luo2023a}. The calculated abundances of CO toward specific sightlines could also be biased by the method of deriving the column densities, excitation, and optical depths. 

The abundance of molecular oxygen (O$_2$) from both observations and chemical models is negligible compared with atomic oxygen in dense clouds \citep{Larsson2007,Goldsmith2011,Wakelam2019}. Thus, the majority of the oxygen element should be in the atomic form in dense clouds. The measured mean value of atomic oxygen abundance (with respect to total H) in the moderate density cloud ($n_{\rm H_2}$ = a few 10$^4$\,cm$^{-3}$) is (2.51$\pm$0.69)$\times$10$^{-4}$ \citep{Lis2023}. 

Since there are no observable rotational or vibrational transitions of N$_2$ in dense clouds, the main reservoir of nitrogen in molecular clouds is still under debate \citep{Womack1992,Maret2006,Furuya2018}. Measurements from N$_2$H$^+$ in different environments suggest that N$_2$ should be abundant in dense regions \citep{Womack1992,Furuya2018}. 

To examine the uncertainties of CRIR induced by CO and O abundances, and whether the assumed abundances of N$_2$ are reasonable, we calculate the CRIR for each of the above chemical simulations, with the abundances (with respect to total H, including the uncertainties) of CO, O, and N$_2$ to be $(1.0\pm0.3)\times10^{-4}$, (2.51$\pm$0.69)$\times$10$^{-4}$, and $(3.8\pm0.5)\times10^{-5}$.

\begin{figure*}
	\includegraphics[width=2.0\columnwidth]{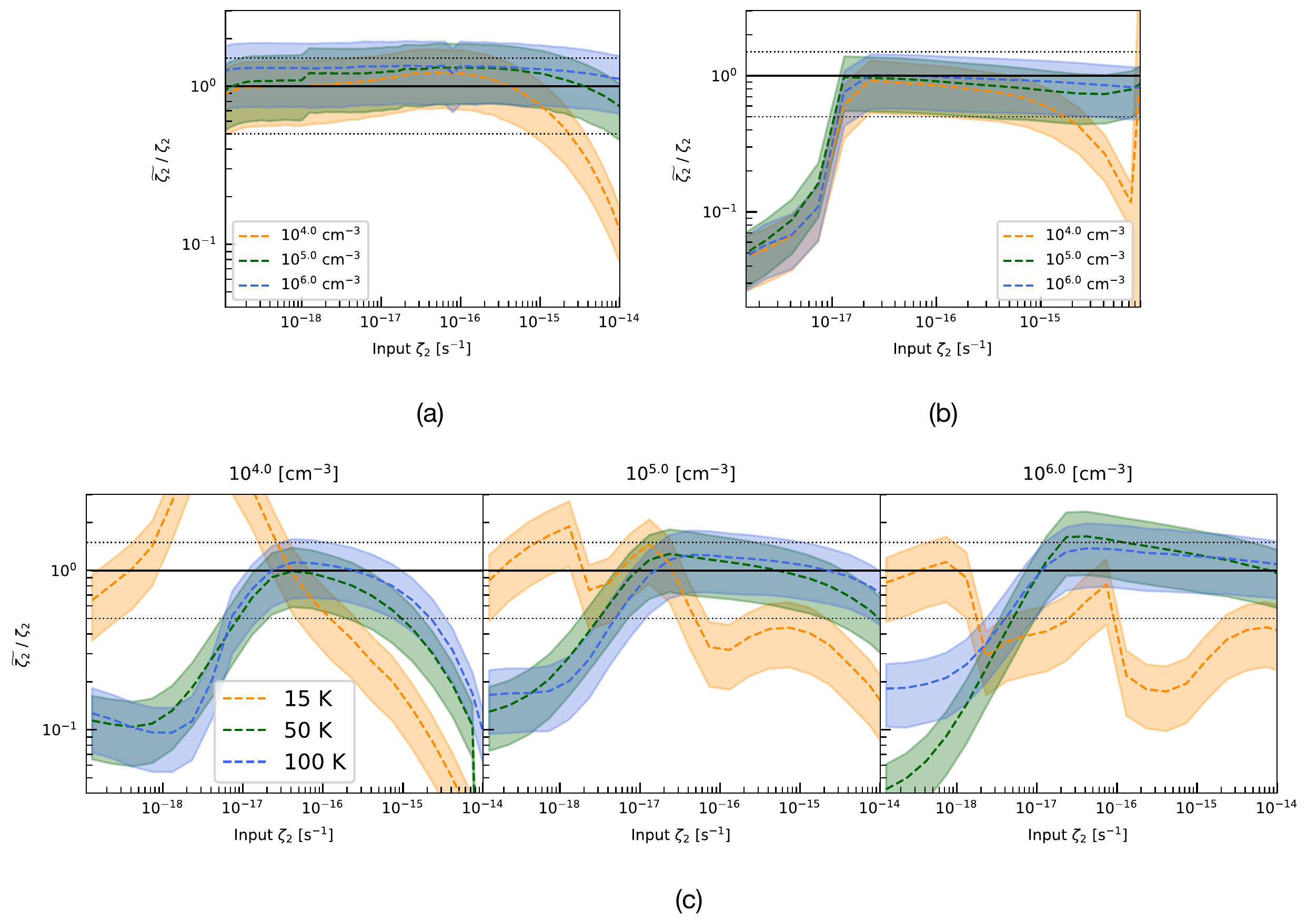}
    \caption{Comparison of $\widetilde{\zeta_2}/\zeta_2$ and uncertainties induced by gas phase abundances of CO, O, and N$_2$ for (a) {\sc 3d-pdr} simulations, (b) HMC simulations from {\sc uclchem}, and (c) isothermal cloud simulations from {\sc uclchem}. The dashed curves and the shadow regions represent the mean value and 1$\sigma$ deviation from our calculation. Orange, green, and blue lines represent different densities (in (a) and (b)) or temperatures (in (c)). Dashed horizontal lines represent 50\% deviation from the input $\zeta_2$.}
    \label{fig:fig8}
\end{figure*}

Figure \ref{fig:fig8} shows the mean value (dashed lines) and 1$\sigma$ deviation (shadow regions) of our analytic approach with comparison to the input $\zeta_2$ from {\sc 3d-pdr}, isothermal simulations from {\sc uclchem}, and HMC simulations from {\sc uclchem}, respectively. 
As can be seen, the derived mean values of CRIR are mostly within 50\% deviation for {\sc 3d-pdr} and HMC simulations with {\sc uclchem} even if we consider the uncertainties induced by CO, O, and N, and the results in Sect.~\ref{sec:results} remain the same. 

Larger differences can be found in isothermal simulations with {\sc uclchem}. For high-temperature models ($T_{\rm k} = 50, 100$ K) with $\zeta_2$ between 10$^{-17}$ to 10$^{-14}$\,s$^{-1}$, the mean value is consistent with that of Fig.~\ref{fig:fig4} and the deviation induced by the above uncertainties are mostly within 50\%. When $\zeta_2$ below a few 10$^{-18}$\,s$^{-1}$, the deviation induced by the above uncertainties would tend to decrease the calculated CRIR by up to a factor of $\sim$10. The results suggest that the assumed initial abundances of O or N$_2$ deviate from the simulations at low CRIR. For low-temperature models ($T_{\rm k} = 15$ K), the predicted mean value is significantly different from the previous results in Fig.~\ref{fig:fig4}, suggesting an unpredictable behavior at the low temperature and the analytic approach should not be used in such environments. In the following analysis, we consider all the above uncertainties when calculating the CRIR using Eq.~\ref{eq:eq8}.

\subsection{Comparison between our calculations and the literature}\label{sec:compare}

We apply our method to some well-known sources with reported CRIR values found in the literature.

\subsubsection{OMC-2 FIR\,4}\label{sec:fir4}

At a distance of $\sim$420\,pc \citep{Hirota2007}, OMC-2 FIR\,4 is located in the north of the Orion KL object, which hosts a few embedded low- and intermediate-mass protostars \citep{Shimajiri2008}. Multi-transitions of N$_2$H$^+$ ($J$=6-5 to 12-11), HCO$^+$ ($J$=6-5 to 13-12), and H$^{13}$CO$^+$ ($J$=6-5 to 7-6) have been observed toward this source with {\it Herschel}. The detailed spectral line energy distribution (SLED) analysis can be found in \citet{Ceccarelli2014b}.

The non-local thermal equilibrium (LTE) results for the warm core component and the envelope are taken from Table 1 in \citet{Ceccarelli2014b}. 
The column density of H$_2$ ($N_{\rm H_2}$) has relatively larger uncertainties than those of molecular lines. The value of $N_{\rm H_2}$ derived from the modeled source size ($\sim$1600\,AU) in \citet{Ceccarelli2014b} is 9.6$\times$10$^{23}$\,cm$^{-2}$. However, this value is higher than that given by radiative transfer modeling of dust spectral energy distribution \citep[SED, $N_{\rm H_2}$ = 2.2$\times$10$^{23}$\,cm$^{-2}$,][]{Crimier2009} and that measured from 3\,mm continuum \citep[1--9$\times$10$^{23}$\,cm$^{-2}$,][]{Fontani2017}. In regards to the 3\,mm continuum, which could be contributed by both free-free emission and dust thermal emission, the derived $N_{\rm H_2}$ should be considered as an upper limit. 

On the other hand, due to the lack of velocity information in the dust continuum, the core and envelope components cannot be distinguished. 
To simplify our calculations, a median value of $N_{\rm H_2}$ = (5$\pm$1.3)$\times$10$^{23}$\,cm$^{-2}$ is adopted for the warm core component (which is the dominant component). The minimum and maximum values of $N_{\rm H_2}$ adopted are 1$\times$10$^{23}$\,cm$^{-2}$ and 9$\times$10$^{23}$\,cm$^{-2}$, respectively. For the envelope component, the value of $N_{\rm H_2}$ = 6.6$\times$10$^{22}$\,cm$^{-2}$ is adopted according to the models presented in \citet{Ceccarelli2014b}.

Following Eq.~\ref{eq:eq8}, the derived CRIR toward the warm core component is $\zeta_2$ = (1.0$\pm$0.3)$\times$10$^{-14}$\,s$^{-1}$, with a lower limit of (0.6$\pm$0.2)$\times$10$^{-14}$\,s$^{-1}$ (adopting $N_{\rm H_2}$ = 9$\times$10$^{23}$\,cm$^{-2}$) and an upper limit of (5.2$\pm$1.4)$\times$10$^{-14}$\,s$^{-1}$ (adopting $N_{\rm H_2}$ = 1$\times$10$^{23}$\,cm$^{-2}$). 
The models of \citet[][]{Ceccarelli2014b} result in a CRIR more than two orders of magnitude higher than ours (they find $\zeta_2$ $\sim6\times10^{-12}\,{\rm s}^{-1}$). We argue that this may be related to their overestimation of N$_2$H$^+$ and HCO$^+$ abundances by 2$\sim$3 orders of magnitude. According to the non-LTE analysis in \citet[][]{Ceccarelli2014b}, the calculated abundances of HCO$^+$ and N$_2$H$^+$ should be 7.3$\times$10$^{-11}$ and 3.1$\times$10$^{-11}$, respectively. However, the resultant abundances of HCO$^+$ and N$_2$H$^+$ from their chemistry analysis are 10$^{-7}$ and 3$\times$10$^{-8}$.

The derived CRIR toward the envelope is $\zeta_2$ = (1.0$\pm$0.3)$\times$10$^{-14}$\,s$^{-1}$, which is slightly lower than the value given by \citet[][$\zeta_2$ = 1.5--8$\times$10$^{-14}$\,s$^{-1}$]{Ceccarelli2014b} and recent chemical modeling from carbon-chain molecules \citep[$\zeta_2$ $\approx$ 4$\times$10$^{-14}$\,s$^{-1}$,][]{Fontani2017,Favre2018}. 

\subsubsection{L1157-B1}\label{sec:l1157-b1}

L1157-B1 is a well-known line-rich shock region, which is located at the blue-shifted outflow lobe of a low-mass protostar L1157  \citep{Bachiller1997,Lefloch2017,Holdship2019,Codella2020}. Various molecular line emissions (e.g., CH$_3$OH, SiO) have been detected toward L1157-B1 with broad linewidth (a few to $\sim$20\,km\,s$^{-1}$) and enhanced abundance respect to H$_2$. High sensitivity IRAM 30-m observations found broad linewidth ($4.3\pm0.2$km\,s$^{-1}$) emission of N$_2$H$^+$, while the calculated abundance of N$_2$H$^+$ has large uncertainties due to the unknown kinetic temperature \citep{Codella2013,Podio2014}. 

Recent NH$_3$ observations by JVLA have revealed the distribution of $T_{\rm k}$ at the blue-shifted lobe of L1157 \citep{Feng2022}. In the B1 regions, $T_{\rm k}$ is in the range of 80$-$120 K. 
We take an average $T_{\rm k}$ to be 100 K within the 26$''$ IRAM beamsize toward L1157-B1, and a consistent $n_{\rm H_2}$ of 10$^5$ cm$^{-3}$ as in \citet{Podio2014}. We use the non-local thermal equilibrium (LTE) radiative transfer model {\sc radex} \citep{van2007} to calculate the column density. The collisional excitation rate coefficients of N$_2$H$^+$ from \citet{Daniel2005} are adopted and obtained from the Leiden Atomic and Molecular Database \citep[LAMDA,][]{Schoier2005}. The derived column density of N$_2$H$^+$ is $(9.2\pm1.0)\times10^{11}$ cm$^{-2}$, and the resultant abundance is $(9.2\pm1.0)\times10^{-10}$. 
The abundances of CO and HCO$^+$ are $1\times10^{-4}$ and $7\times10^{-9}$, respectively \citep{Podio2014}. The derived CRIR toward L1157-B1 from Eq.~\ref{eq:eq8} is $\zeta_2 = (2.2\pm0.4)\times10^{-16}$ s$^{-1}$, which is in good agreement with the value from chemical modeling \citep[$\zeta_2 \approx 3\times10^{-16}$ s$^{-1}$,][]{Podio2014,Benedettini2021}.

\section{The CRIR in HMSFRs}\label{sec:crir in msfr}

To statistically investigate how the CRIR varies in different stages of HMSFRs, we calculate CRIR using our method toward a total of 82 samples from \citet{Purcell2006,Purcell2009}, \citet{Gerner2014,Gerner2015}, and \citet{Entekhabi2022}. These samples include 20 HMPOs, 54 HMCs, and 8 ultra-compact H\,{\sc ii} (UCH\,{\sc ii}) regions.

In the hot core samples of \citet{Purcell2006,Purcell2009}, the gas temperatures are in the range of $28\pm7$ to $131\pm53$ K, which are derived from the rotational diagram of CH$_3$CN. The column densities of HCO$^+$ are derived from the optically thin H$^{13}$CO$^+$ emission, assuming $^{12}$C/$^{13}$C =50 and $T_{\rm ex} = 50$\,K. The column densities of N$_2$H$^+$ are derived from the hyperfine transitions with the assumption of $T_{\rm ex} = 10$\,K and are corrected for the beam dilution with dust continuum. The column and volume densities of H$_2$ are derived from 1.2\,mm continuum emission (see \citet{Purcell2009} for details). There are no observations of CO isotopologues, in our calculation, a constant abundance of CO ($1.0\pm0.3\times10^{-4}$) is assumed for all the hot core samples from \citet{Purcell2006,Purcell2009}. 

In the samples of \citet{Gerner2014,Gerner2015}, the gas temperatures of HMPOs, HMCs, and UCH\,{\sc ii} are 29.5, 40.2, and 36\,K, respectively. The column densities of HCO$^+$ are derived from two transitions of H$^{13}$CO$^+$ (1--0 and 3--2) using {\sc radex}, assuming $^{12}$C/$^{13}$C = 89\footnote{Although the adopted value is higher than the local ISM \citep[$\sim$65,][]{Milam2005} and those from \citet{Purcell2006,Purcell2009}, a variance of  $^{12}$C/$^{13}$C by 30\% would only result in an uncertainty of $\zeta_2$ by $\sim$1\%.}. The column densities of N$_2$H$^+$ are calculated with the hyperfine structure fitting routine. The column densities of CO are derived from C$^{18}$O with LTE, assuming $^{16}$O/$^{18}$O = 500. The H$_2$ column densities are derived from dust emission at 850\,$\mu$m with JCMT or 1.2\,mm with IRAM-30m (see \citet{Gerner2014} for details). The volume densities of H$_2$ are derived by assuming a typical source size of 0.5\,pc for all sources. We note that the choice of source size is based on the IRAM-30 m beam size, which may underestimate the actual physical size of these systems. Thus, the derived CRIR may also be underestimated.

\begin{figure}
	\includegraphics[width=1.0\columnwidth]{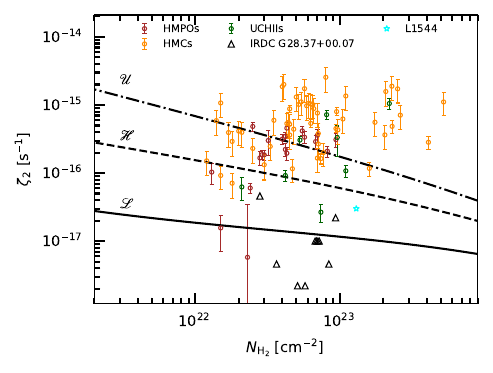}
    \caption{The calculated $\zeta_2$ toward HMPOs (brown), HMCs (orange), and UCH\,{\sc ii}s (green). Black triangles represent CRIR from IRDC G28.37+00.07, which are taken from \citet{Entekhabi2022}. The cyan star represents CRIR from prestellar core L1544, which is taken from \citet{Redaelli2021}. The black solid, dashed, and dash-dotted curves denote the theoretical CR attenuation models $\mathscr{L}$, $\mathscr{H}$, and $\mathscr{U}$ by \citet{Padovani2024}.}
    \label{fig:fig10}
\end{figure}

Figure \ref{fig:fig10} shows the calculated $\zeta_2$ from the above samples. For comparison, the CRIR values from L1544 \citep{Redaelli2021} and IRDC G28.37+00.07 \citep{Entekhabi2022}, and the theoretical CR attenuation models $\mathscr{L}$, $\mathscr{H}$, and $\mathscr{U}$ from \citet{Padovani2024} are also overlaid. The models $\mathscr{L}$, $\mathscr{H}$, and $\mathscr{U}$ are based on different assumptions on the low-energy slope of the interstellar CR proton spectrum (see \citet{Ivlev2015} and \citet{Padovani2024} for a detailed description of the parameterization of the CR spectrum), which have energy density $\varepsilon_{\rm CR} \approx$ 0.65, 1.18 and 2.28 eV cm$^{-3}$, respectively. The trends of the CRIR in Fig. \ref{fig:fig10} also account for the ionization due to primary and secondary electrons. It is noted that all the physical parameters ($T_{\rm kin} \geq 30$ K, $n_{\rm H_2} > 10^4$ cm$^{-3}$, no CO depletion) are within the allowable parameter space that the analytic approach gives a good approximation.
The average values of $\zeta_2$ in HMPOs, HMCs, and UCH\,{\sc ii} are (2.5$\pm$1.3)$\times$10$^{-16}$, (7.4$\pm$5.0)$\times$10$^{-16}$, and (3.4$\pm$3.4)$\times$10$^{-16}$\,s$^{-1}$, respectively. The values in these targets are over an order of magnitude higher than the measurements in prestellar cores (e.g., L1544 and G28.37+00.07) and that predicted by the model $\mathscr{H}$, where the latter describes the average decrease of the observationally estimated $\zeta_2$ \citep{Padovani2022,Padovani2023a}.

Interestingly, observations from $Fermi$ and LHAASO also find the ubiquitous $\gamma$-ray emissions in HMSFRs (e.g., W43) from sub-GeV to PeV \citep{Yang2018,Yang2020,Vink2022,delaFuente2023}, suggesting an efficient acceleration of particles in active star-forming environments \citep{Bykov2020,Tibaldo2021,Owen2023}. While the high energy ($E \gg 10$ GeV) $\gamma$-ray could originate from CRs accelerated by stellar wind of massive stars \citep{Peron2024}, the lower energy $\gamma$-ray could come from the embedded massive protostars which undergo their accretion phase and in which the first-order Fermi acceleration mechanism operates, e.g., jet shocks \citep{Padovani2015,Padovani2016,Lattanzi2023}, shocks on protostellar surfaces \citep{Padovani2015,Padovani2016,Gaches2018}, H\,{\sc ii} regions \citep{Padovani2019,Meng2019}, and wind-driven shocks \citep{Bhadra2022}. These processes would accelerate CRs to tens of GeV (depending on the actual physical conditions), leading to a detectable $\gamma$-ray emission toward individual protostar \citep{Araudo2007,Bosch-Ramon2010,Padovani2015,Padovani2016}. The detection of $\gamma$-ray emission at $\sim$1 GeV from the protostellar jet in HH 80-81 and $\sim$10 GeV level toward a young massive protostar supports these scenarios \citep{Yan2022,Wilhelmi2023}.

The high values of $\zeta_2$ from our work are also consistent with the model prediction under the existence of internal CR source models as presented by \citet{Gaches2019}, in which the CRIR is in the range of 10$^{-16}$\,s$^{-1}$ to 10$^{-14}$\,s$^{-1}$ at a column density of $\sim$10$^{23}$\,cm$^{-2}$. In addition, observations toward B335 also suggest high CRIR near the central protostars \citep{Cabedo2023}. 
The increasing trend of CRIR from HMPOs to HMCs is consistent with our understanding that HMCs are in a more evolved phase and that star-formation activities (e.g., shocks) in HMCs may have stronger impacts on the surrounding gas. At the late stage of HMSFRs, once the central massive protostars have gathered enough mass and evolved to the zero age main sequence, the stellar wind and radiation feedback would drive the surrounding gas and lead to an expansion of UCH\,{\sc ii} regions. And because the accretion and outflows/jets gradually disappear in the late stage, the CRIR could be lower. 

Such a methodology could be, in principle, used to estimate CRIR of external galaxies, where the physical properties could be quite different to those in the Milky Way \citep{Rosolowsky2021}. However, we should note that a better understanding of the source properties (e.g., density, metallicity) and a correction for the beam dilution are needed before using such a method on distant starburst galaxies. This is because, the spatial resolution of distant galaxies is usually quite low, thus, the average volume density within the beam is not supposed to be high enough to give a good approximation at very high CRIR ($\zeta_2 \geq 10^{-14}$ s$^{-1}$, see Sect. \ref{sec:results}). For cosmic noon starburst galaxies where the CRIR is supposed to be extremely high \citep[e.g., orders of magnitude above $\sim10^{-16} - 10^{-14}$ s$^{-1}$,][]{Indriolo2018}, the analytic approach could only give a lower limit.

\section{Conclusions}\label{sec:conclusions}

We present a new analytic approach to estimate the CRIR using the abundances of HCO$^+$, N$_2$H$^+$, and CO. The basic assumptions of deriving the analytic approach are: 1) HCO$^+$, N$_2$H$^+$, H$_3^+$, and H$_3$O$^+$ are the main ions, and 2) warm gas environments (depletion of molecules can be ignored). 
Our analytic approach has been tested with different {\sc 3d-pdr} and {\sc uclchem} simulations.
The main conclusions are as follows:
\begin{enumerate}
    \item The analytic approach gives a reasonable estimation of $\zeta_2$ (to within 50\% deviation) in dense warm gas (e.g., $n_{\rm H} \geq 10^4$\,cm$^{-3}$, $T \geq 50$\,K) for $A_{\rm V} \geq 4$ mag, $t \geq 2\times10^4$ yr. The higher the gas density, the better the estimation from our analytic approach. Such environments may correspond to HMPOs and HMCs.
    \item Application of our method to the OMC-2 FIR\,4 envelope and L1157-B1 results in $\zeta_2$ values of $(1.0\pm0.3)\times10^{-14}$ s$^{-1}$ and $(2.2\pm0.4)\times10^{-16}$ s$^{-1}$, which are in good agreement with the estimation of $\zeta_2$ from the chemical modeling in the literature, suggesting wide applicability of our method in warm gas environments. 
    \item The values of $\zeta_2$ toward a total of 82 samples of HMPOs, HMCs, and UCH\,{\sc ii} have been derived with our analytic approach. The average value of $\zeta_2$ in HMCs is over an order of magnitude higher than the prediction of canonical CR attenuation models. Our results suggest that the internal massive protostars are likely to be responsible for the high CRIR.
\end{enumerate}

Future high spatial resolution interferometry observations with multi-transitions of molecular lines will allow the more robust estimation and direct mapping of CRIR in HMSFRs, which is helpful to understand the origin of the high CRIR \citep{Padovani2019,Padovani2021}.

\begin{acknowledgements}
We thank Evelyne Roueff for her careful review of the chemical reactions in our manuscript, particularly for pointing out that the endothermic reaction can only occur under specific conditions. BALG acknowledges support from a Chalmers Cosmic Origins postdoctoral fellowship. 
\end{acknowledgements}

%
%

\bibliographystyle{aa}

\bibliography{reference} 



\begin{appendix} 

\section{Comparison of electron fraction from our assumption and model prediction}\label{sec:a1}

\begin{figure*}
	\includegraphics[width=2.0\columnwidth]{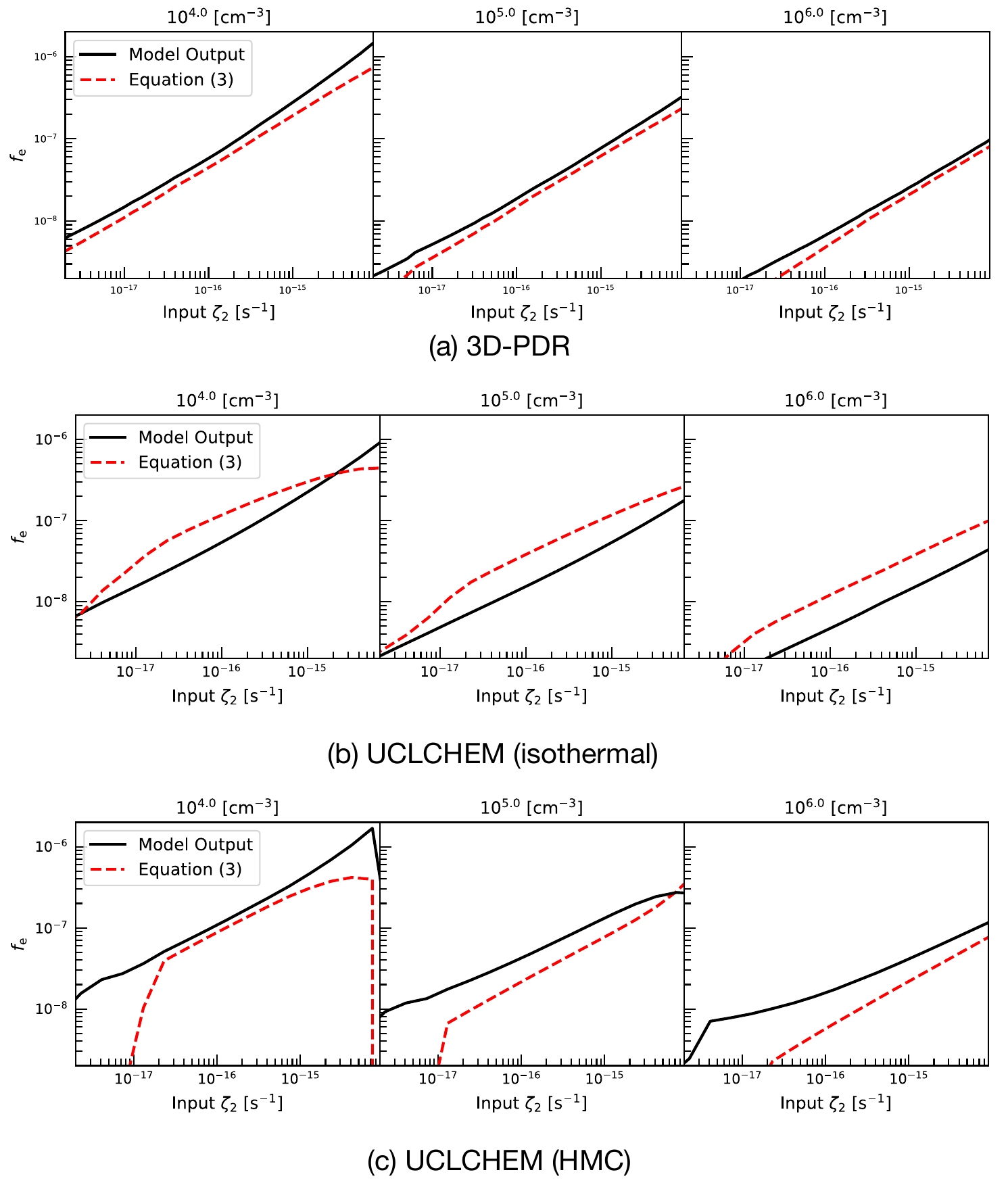}
    \caption{(a): The electron fraction as a function of $\zeta_2$ as obtained from the {\sc 3d-pdr} output (black solid curves) and from Eq.~\ref{eq:eq4} (red dashed curves). (b): The same as (a) while for the isothermal simulations from {\sc uclchem}. (c): The same as (a) while for the HMC simulations from {\sc uclchem}.}
    \label{fig:a1}
\end{figure*}

Figure \ref{fig:a1} shows the comparison between the electron fraction ({\it f}(e$^-$)) obtained from model simulations and that of Eq.~\ref{eq:eq4}. As can be seen, our assumption in Eq.~\ref{eq:eq4} gives reasonable results for both {\sc 3d-pdr} and {\sc uclchem} simulations. The deviation between Eq.~\ref{eq:eq4} and model prediction is within a factor of $\sim$2 for all {\sc 3d-pdr} simulations and isothermal simulations from {\sc uclchem}. For the HMC simulations from {\sc uclchem}, the deviation is within a factor of 4 if $10^{-17} \leq \zeta_2/{\rm s}^{-1} \leq 10^{-14}$. For $\zeta_2 < 10^{-17}$\,s$^{-1}$, Eq.~\ref{eq:eq4} underestimate the electron fraction since we did not consider the contribution of other ions (e.g., C$^+$, Mg$^+$). 

\section{The reaction rates that appears in the analytic approach}\label{sec:rates}

Table \ref{table:rates} lists the reaction rates that appear in Sect.\ref{sec:methods}, where the rates are taken from UMIST database \citep{Millar2024}.

\begin{table*}
\caption{The reaction rates of reactions in Sect. \ref{sec:methods}.}             
\label{table:rates}      
\centering          
\begin{tabular}{c c | c c }     
\hline\hline       
Reactions & Rates & Reactions & Rates\\
\hline
 & cm$^{-3}$ s$^{-1}$ & & cm$^{-3}$ s$^{-1}$ \\
\hline                    
\ce{H2 + CR -> H2+ + e^-} & 0.88$\zeta_2$ & \ce{H2+ + H2 -> H3+ + H} & $2.08\times10^{-9}$ \\
\ce{H3+ + CO -> HCO+ + H2} & $1.36\times10^{-9}(\frac{T}{300})^{-0.14}e^{\frac{3.4}{T}}$ & \ce{H3+ + CO -> HOC+ + H2} & $8.49\times10^{-10}(\frac{T}{300})^{0.07}e^{\frac{-5.2}{T}}$ \\
\ce{H3+ + N2 -> N2H+ + H2} & $1.8\times10^{-9}$ & \ce{H3+ + O -> OH+ + H2} & $4.65\times10^{-10}(\frac{T}{300})^{-0.14}e^{\frac{-0.67}{T}}$ \\
\ce{H3+ + O -> H2O+ + H} & $2.08\times10^{-10}(\frac{T}{300})^{-0.4}e^{\frac{-4.86}{T}}$ & \ce{H3+ + e^- -> H2 + H} & $2.34\times10^{-8}(\frac{T}{300})^{-0.52}$ \\
\ce{H3+ + e^- -> H + H + H} & $4.36\times10^{-8}(\frac{T}{300})^{-0.52}$ & \ce{HCO+ + e^- -> CO + H} & $2.4\times10^{-7}(\frac{T}{300})^{-0.69}$ \\
\ce{N2H+ + CO -> HCO+ + N2} & $8.8\times10^{-10}$ & \ce{N2H+ + e^- -> H + N2} & $3.06\times10^{-7}(\frac{T}{300})^{-0.06}$ \\
\ce{N2H+ + O -> OH+ + N2} & $1.4\times10^{-10}$ & \ce{OH+ + H2 -> H2O+ + H} & $1.27\times10^{-9}(\frac{T}{300})^{0.18}$ \\
\ce{H2O+ + H2 -> H3O+ + H} & $9.7\times10^{-10}$ & \ce{H3O+ + e^- -> OH + H + H} & $3.05\times10^{-7}(\frac{T}{300})^{-0.5}$ \\
\ce{H3O+ + e^- -> OH + H2} & $5.37\times10^{-8}(\frac{T}{300})^{-0.5}$ & \ce{H3O+ + e^- -> H2O + H} & $7.09\times10^{-8}(\frac{T}{300})^{-0.5}$ \\
\hline                  
\end{tabular}
\end{table*}


\end{appendix}
%
%
\end{document}